# Edge Dominating Capability based Backbone Construction in Wireless Networks

Congcong Chen[1], Jiguo Yu[1,2], Xiujuan Zhang[1]

[1] School of Computer Science, Qufu Normal University, Rizhao, Shandong, 276826, China

[2] Key Laboratory for Intelligent Control Technique of Shandong Province, Rizhao, Shandong, 276826, China

**Abstract**

Constructing a connected dominating set as the virtual backbone plays an important role in wireless networks. In this paper, we propose two novel approximate algorithms for dominating set and connected dominating set in wireless networks, respectively. Both of the algorithms are based on edge dominating capability which is a novel notion proposed in this paper. Simulations show that each of proposed algorithm has good performance especially in dense wireless networks .

## 1. Introduction

Wireless networks such as ad-hoc networks and sensor networks consist of a number of wireless autonomous nodes, which communicate through wireless radio technology. They do not rely on any existing or predefined network infrastructure. Wireless networks are widely deployed for many applications such as automated battlefield operations, disaster rescues, environmental detections and so on. However, due to un-predefined infrastructure, the transmission of information between the nodes and other related routing tasks are much more complex than that of wired networks. A virtual backbone network is came up to overcome this shortcoming, non-backbone nodes communicate through the backbone nodes. This greatly reduces the routing search space, and can effectively implement unicast, multicast and fault-tolerant routing. Clustering based on dominating set is an important method to construct a virtual backbone network in wireless networks [1]. A dominating set (DS) of a graph $G = (V, E)$ is a node subset $S \subseteq V$, such that every node $v \in V$ is either in $S$ or adjacent to a node of $S$. A node in $S$ is said to dominate itself and all adjacent nodes. We can use the nodes in a dominating set as cluster-heads and assign each node to a cluster corresponding to a node that dominates it. Only the nodes in the dominating set communicate directly, other nodes communicate through their neighborhood dominators. In general, one wishes to find a small number of clusterheads. That is, a small dominating set, in order to simplify the network structure as much as possible.

According to connectivity, dominating set can be classified into Independent Dominating Sets (IDSs), Connected Dominating Sets (CDSs) and Weakly Connected Dominating Sets (WCDSs) [2]. An IDS is a dominating set $S$ of a graph $G$ in which there are no adjacent nodes. Fig. 1(a) shows a sample independent dominating set where black nodes form an IDS of the graph. A CDS is a subset $S$ of a graph $G$ such that $S$ forms a dominating set and $G[S]$ is connected, where $G[S]$ is the induced subgraph. Fig. 1(b) shows a sample CDS. If the message routes along a CDS, most of the redundant broadcasts can be eliminated [3]. A weakly induced subgraph $G<S>$ is a subgraph of a graph $G$ that contains the nodes of $S$, their neighbors and all edges of the original graph $G$ with at least one endpoint in $S$. A subset $S$ is a weakly connected dominating set, if $S$ is dominating and $G<S>$ is connected [4]. Black nodes in Fig. 1(c) form a WCDS. In [5], the author proposed the notion of extended dominating sets (EDSs). A subset $S$ is an EDS if every node is (a) in $S$, (b) a regular neighbor of a node in $S$, or (c) a quasi neighbor of $k$ nodes in $S$. An edge dominating set of a graph $G$ is a subset $M \subseteq E$ such that each edge in $E$ shares an endpoint with some edges in $M$ [6]. The Minimum Edge Dominating Set problem asks to find an edge dominating set of minimum cardinality $|M|$. It has been proved to be NP-complete in [7].

In this paper, we consider the properties of edge and node in an undirected graph. According to the dominant capability between edge and node, we propose the notion of edge dominating capability (EDC), and design two approximate algorithms based on EDC for a DS and a CDS, respectively.

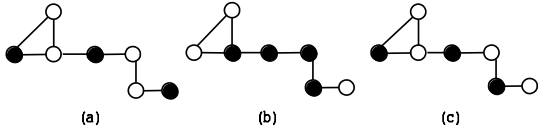

Fig. 1 (a)IDS (b)CDS (c)WCDS.

The rest of this paper is organized as follows. Section 2 reviews the existing algorithms for constructing DS and CDS. Section 3 proposes the notion of edge dominating capability. In section 4, we design two approximate algorithms to construct small DSs and CDSs, respectively. And we evaluate the performance of our algorithms by theoretical analysis. Section 5 shows the results of simulation. Finally, section 6 concludes this paper.

## 2. Related Work

In this section, we review some existing algorithms for constructing a dominating set (DS) and a connected dominating set (CDS).

Finding a small dominating set is one of the most fundamental problems in traditional graph theory, and belongs to be NP-hard in general, but efficient approximate algorithms do exist. In [8], the authors gave a simple greedy algorithm for finding small dominating sets in undirected graphs of $n$ nodes and $m$ edges, and showed that $d_g \leq N + 1 - \sqrt{2M+1}$, where $d_g$ is the cardinality of the dominating set returned by the algorithm. Kuhn et al. presented a new fully distributed approximate algorithm based on LP relaxation techniques [9]. For an arbitrary parameter $k$ and maximum degree $\Delta$, the algorithm computes a dominating set of expected size $O(k \Delta^{\frac{2}{k}} \log \Delta |DS_{OPT}|)$ in $O(k^2)$ rounds where each node has to send $O(k^2 \Delta)$ messages of size $O(\log \Delta)$. This is the first algorithm which achieves a non-trivial approximate ratio in a constant number of rounds. In [10], authors gave a simple and efficient distributed algorithm for constructing minimal dominating set in wireless sensor networks. The dominating set constructed by the proposed algorithm can be adaptive to the changes of network topology. In [11], authors analyzed edge dominating set from a parameterized perspective and proved that this problem is in $FPT$ for general graphs. Authors proposed efficient enumeration-based exact algorithms for finding an edge dominating set.

There exist abundant CDS formation protocols for wireless networks in the literature. Based on their efficiency in terms of forming a small CDS and overhead in terms of messages and time complexity, these protocols can be classified into four categories in [12]: global, quasi-global, quasi-local, and local. Here we introduce some important CDS constructing algorithms. Wu and Li proposed a completely local algorithm where each node knows the connectivity information within the 2-hop neighborhood [12]. The generated CDS is easy to be maintained. But the size of the CDS is large. Thus they gave two rules to prune the generated CDS [13]. Wu et al. proposed a general framework of the iterative local solution (ILS) for computing a connected dominating set in ad hoc wireless networks [5]. This approach uses an iterative application of a selected local solution. Each application of the local solution enhances the result obtained from the previous iteration, but each is based on a different node priority scheme. In [14], the authors proposed a distributed algorithm for CDS in an UDG. This algorithm consists of two phases and has a constant approximation ratio of 8. The algorithm fist constructs a spanning tree. Then each node in a tree is examined to find an MIS for the first phase, more nodes are added to connect those black nodes. In [15], Qayyum et al. proposed an efficient broadcast scheme called multi-point relying (MPR). In MPR, each host designates a small set of 1-hop neighbors to cover its 2-hop neighbors. In [16], the authors proposed an algorithm which is based on nodes neighborhood to construct a minimum connected dominating set in wireless networks. The time complexity and message complexity are O(n log n) and O(n), respectively. Regarding to the topological changes due to power constraints, authors represented a repair algorithm that reconstructs the MCDS. In [17], Li et al. investigated the problem of constructing quality CDS in terms of size, diameter, and average backbone path length, and proposed two centralized algorithms having constant performance ratios for its size and diameter of the constructed CDS.

Han and Jia proposed an area-based distributed algorithm for WCDS construction in wireless ad hoc networks [18]. This algorithm has both time and message complexity of $O(n)$, the size of WCDS constructed is within a constant approximation ratio. Chen and Liestman also proposed a region-based algorithm [19]. In this approach, they divided the graph into regions, and the partitioning phase is partly based on a Minimum Spanning Tree algorithm. The size of regions is controlled by picking a value $x$. They also presented two centralized algorithms and one distributed algorithm for finding a small WCDS, these algorithms are all based on the idea of piece [20]. Wu et al. proposed four algorithms for constructing a small extended dominating set [21]. More recently, Yu et al. proposed four novel algorithms to construct extended weakly connected dominating sets [22].

## 3. Edge Dominating Capability

In this section, we describe the concept of edge dominating capability (EDC) based on the dominant property between edge and node in an undirected graph. EDC is the main consideration when we construct a dominating set and a connected dominating set in this paper.

We make the following assumption. In an undirected graph $G = (V, E)$, if the degree of a node $v$ is $d_v$, then each of the $d_v$ edges can dominates the node $v$, and the dominated probability of the node $v$ is $1/d_v$. That is, an edge fractionally dominates its endpoints. We define the dominating capability of an edge is the sum of dominated probabilities of the two endpoints incident with the edge. As shown in Fig. 2, the degree of node 1 is 1, the degree of node 2 is 3, thus according to our assumption, the dominant capability of the edge (1,2) to node 1 is 1, the dominant capability to node 2 is $1/3$. Therefore, the weight of (1,2) is $4/3$.

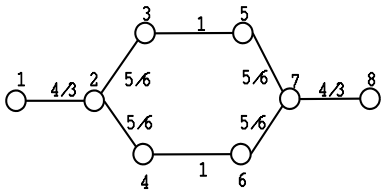

Fig. 2  The weight of each edge.

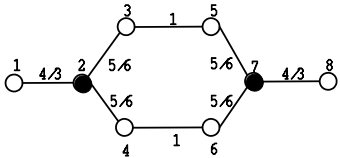

Fig. 3  The set of black nodes {2,7} forms a DS.

According to the characteristic (such as edge dominating capability) of edge, a dominating set can be selected. In a real network, we can select the nodes in a path with largest weight as the dominators based on the idea of edge dominating capability, to forward information rather than to broadcast the information. This can greatly reduce the energy consumption and information redundancy, and quickly transfer information to destination. Thus, our algorithm essentially depends on the dominating capability of the edge to node, to find a dominating set in a wireless network.

## 4. Approximate Algorithms

In this section, we present two approximate algorithms to construct a small dominating set and connected dominating set in a wireless network, respectively. And we evaluate the performance of our algorithms by theoretical analysis.

### 4.1 An approximate algorithm for DS

According to the assumption of section 3, we definite that each edge of a graph can dominate its two endpoints fractionally, and its weight is the sum of the edge to its endpoints' dominated probabilities. Now, we present an approximate algorithm base on EDC to find a small dominating set in a wireless network. The algorithm is called as EDC-DS algorithm.

EDC-DS algorithm:
1. An edge with maximum weight (edge dominating capability) is selected as a dominant edge. If more than one edges have maximum weights, we select all the edges as dominant edges at the same time.

2. A node which is dominated by a dominant edge with minimum dominant capability is selected as a dominator. That is, we select the node $v$ which connects a dominant edge as a dominator, when the dominant capability $1/d_v$ is minimum. If the dominant capabilities of two endpoints are same, then we select the node with the minimum $id$ as a dominator.

3. If the selected dominant edge has had one dominator endpoint, or its two endpoints have already been dominated, then it is not necessary to select an endpoint as a dominator.

4. The same procedure is implemented iteratively, until each node in the network is dominated by at least one dominator.

5. The set of selected dominators forms a dominating set of the network.

Fig. 4 shows the whole implement procedure of finding a dominating set in the original network (as shown in Fig. 2) using EDC-DS algorithm, where black nodes represent the selected dominators, and red edges is the selected dominant edges. In the end, the set of dominators {2,3,4,7} forms a dominating set of the network.

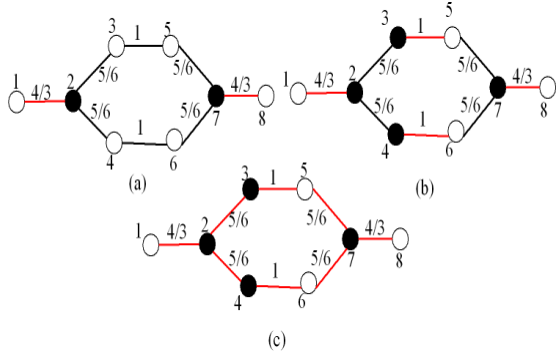

Fig. 4 (a)The first iteration, nodes 2 and 7 are selected as dominators. (b)The second iteration, nodes 3 and 4 are selected ad dominators. (c)The third iteration, each dominant edge has at least one dominator endpoint.3.2 Equations

**Theorem 1.** The set of selected dominators by our algorithm forms a dominating set of a network.

**Proof.** From the above procedure, we can see that each node of the network is dominated by at least one dominator, the algorithm terminates. That is, each node is either selected as a dominator, or fractionally dominated by a dominator. Therefore, the set of dominators forms a dominating set of the network.

**Theorem 2.** The time and message complexity of our algorithm are $O((e+n+\log n)\log_\Delta n)$ and $O((e+3n)\log_\Delta n)$, respectively. Where $n$ is the number of nodes, $e$ is the number of edges, and $\Delta$ is the maximum degree of the network.

**Proof.** Each node broadcasts its degree, and each edge collects its dominant capabilities to the endpoints. The two procedures need $O(e+n)$ time and $O(e+2n)$ message. After obtaining the weight of each edge, edges are sorted in non-ascending order. This step needs $O(\log n)$ time and $O(n)$ message. The number of iteration of our algorithm is at most $O(\log_\Delta n)$. Therefore, the total complexity of time and message are $O((e+n+\log n)\log_\Delta n)$ and $O((e+3n)\log_\Delta n)$, respectively.

**Theorem 3.** The size of the DS found by our algorithm is at most $(\ln(\Delta+1)+1)|opt|$, where $|opt|$ is the size of the minimum DS of a graph, $\Delta$ is the maximum degree.

**Proof.** Let $opt$ is a minimum DS of a graph $G$, and $|opt|$ is the size of $opt$. Each node in $opt$ at most fractionally dominate $\Delta+1$ nodes. Consequently, $G$ can contain at most $n \leq (\Delta+1)|opt|$ nodes. It follows that

$$|opt| \geq \frac{n}{\Delta+1}.$$

In each iteration of the algorithm, we choose one or more edges with the maximum weight into dominant edges if the rules are satisfied. According to the dominant edges, then we can select at least one node to be a dominator, added into DS. When the algorithm terminates, each node of the graph $G$ is dominated by at least one dominator. Let $a_i$ be the number of nodes which are not dominated after the $i$th iteration, and $a_0 = n$. Consider the $(i+1)$th iteration. Since the addition of the non-dominated nodes of $opt$ would dominate all of the remaining $a_i$ non-dominated nodes. There is at least one non-dominated node of $opt$ which would dominate the number of nodes by at least $\left\lceil \frac{a_i}{|opt|} \right\rceil$.

So we have the relation

$$a_{i+1} \leq a_i - \left\lceil \frac{a_i}{|opt|} \right\rceil$$

$$\leq a_i(1 - \frac{1}{|opt|}) \quad (1)$$

Solving it, we get the following bound

$$a_{i+1} \leq a_0(1 - \frac{1}{|opt|})^{i+1} \quad (2)$$

Setting $i+1 = |opt| \cdot \ln \frac{a_0}{|opt|}$, we have

$$a_{i+1} \leq a_0(1 - \frac{1}{|opt|})^{i+1}$$

$$\leq a_0(1 - \frac{1}{|opt|})^{|opt| \cdot \ln \frac{a_0}{|opt|}}$$

$$\leq a_0(\frac{1}{e})^{\ln \frac{a_0}{|opt|}}$$

$$\leq a_0 \cdot \frac{|opt|}{a_0}$$

$$\leq |opt| \quad (3)$$

That is, after $|opt| \cdot \ln \frac{a_0}{|opt|}$ iterations, we only need to pick at most $|opt|$ additional nodes to dominate the remaining nodes. The total number of nodes that we choose is no more than $|opt|$. Thus, the total number of dominators is $|opt| \cdot \ln \frac{a_0}{|opt|} + |opt|$.

Since $|opt| \geq \frac{n}{\Delta+1}$, the solution found by our algorithm has $|DS|$ nodes,

$$|DS| \leq |opt| + |opt| \cdot \ln \frac{a_0}{|opt|}$$
$$\leq |opt|(1 + \ln \frac{a_0}{|opt|})$$
$$\leq |opt|(1 + \ln \frac{n(\Delta+1)}{n}) \quad (n >> 2)$$
$$\leq |opt|(\ln(\Delta+1)+1) \qquad (4)$$

Therefore, the size of the DS found by our algorithm is at most $(\ln(\Delta+1)+1)|opt|$. The approximation ratio of the algorithm is $1+\ln(\Delta+1)$.

From above theorems, we know that our approximate algorithm can obtain a dominating set with a good performance in polynomial time. However, the result is not necessarily minimum. As shown in Fig. 4, the set {2,7} is a smaller dominating set of the network than {2,3,4,7}. In the following, we improve the above algorithm to obtain a minimum dominating set.

Improved method is as follows. After finishing the first two steps of the algorithm, that is, dominators have been selected in the first iteration, we consider all the edges connected to the dominators as dominant edges. Therefore, it is not necessary to compare the weights of the dominant edges. These dominant edges dominate their another endpoints by their fractionally dominant capability. Then, checking whether there still exist any node not dominated by one dominant edge. If not, the procedure ends. Otherwise, the above procedure iteratively implements.

Fig. 5 uses the network shown in Fig. 2, and implements our improved method to find a smaller dominating set. The method implements only once, we can obtain the final result described in Fig. 5, where black nodes represent dominators, and red edges represent dominant edges. The set {2,7} is a smaller dominating set of the network.

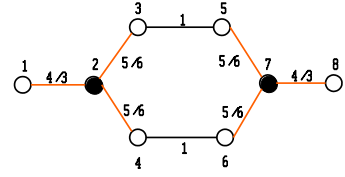

Fig. 5 {2,7} is a smaller DS found by the improved method.

### 4.2 An approximate algorithm for CDS

In this subsection, we present an algorithm called EDC-CDS algorithm to construct a CDS in a wireless network. The proposed algorithm is based on a DS found in the section 4.1, and obtains a small CDS by building a steiner tree. The specific process of connection is as follows.

1. We definite the set $D$ as a minimum DS found in section 4.1, then select the node with minimum $id$ in $D$ as a root.

2. Checking all the nodes of $D$ (except of the root) in an ascending order of $id$ whether there exists a path from each node to the root, and the path only includes the nodes of $D$.

3. If not, select a path which includes the least number of nodes of $V$-$D$ and can not form a loop. And add the nodes of $V$-$D$ in this path as dominators.

4. Else, return to 2.

5. The set of dominators form a CDS called $C$ of the network.

We implement the connection process in the Fig. 5, and obtain a CDS {2,3,5,7} of the graph. The whole process is as shown in Fig. 6, where blue nodes 3 and 5 are the selected connectors.

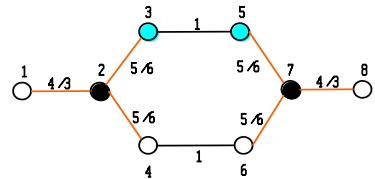

Fig. 6 {2,3,5,7} is a CDS of the graph.

**Theorem 4.** The size of the CDS found by our algorithm is at most $(\ln(\Delta-1)+1)*|opt|$, where $|opt|$ is the size of the minimum CDS of a graph, $\Delta$ is the maximum degree.

**Proof.** Let $u_1$ be an arbitrary node of the $opt$. As $u_1$ is of degree at most $\Delta$, at most $\Delta+1$ distinct nodes can be dominated by $u_1$ (including $u_1$ itself). As $opt$ is a CDS of $G$ with minimum size, there must be another node of $opt$ connected to $u_1$. Let $u_2$ be such a node. Then, at most $\Delta-1$ new distinct nodes are dominated by $u_2$. Again, as $opt$ is a CDS of $G$, there must be another node of $opt$ connecting to either $u_1$ or $u_2$. This node, called $u_3$, dominates at most $\Delta-1$ new distinct nodes. We repeat this argument until we have included all $|opt|$ nodes of $opt$. Thus, $G$ can contain at most $n \leq (\Delta+1)+(\Delta-1)*(|opt|-1)$ nodes. It follows that $|opt| \geq \frac{n-2}{\Delta-1}$.

In each iteration of the algorithm, we select one or more nodes to add into the set $C$, according to the defined rule. Observe that the number of non-dominated nodes is monotonically non-increasing over time. At the beginning of the algorithm, all nodes of the network are non-dominated. As the algorithm implements, the nodes are dominated by dominant edges. When the algorithm terminates, each node is dominated by at least one dominator. Let $a_i$ be the number of nodes which are non-dominated after the $i$th iteration and let $a_0 = n$. Consider the $i+1$th iteration. Since the addition of the non-dominated nodes of $opt$ would dominate all of the remaining $a_i$ non-dominated nodes. There is at least one non-dominated node of $opt$ which would dominate the number of nodes by at least $\left\lceil \frac{a_i}{|opt|} \right\rceil$.

This gives us the recurrence relation,

$$a_{i+1} \leq a_i - \left\lceil \frac{a_i}{|opt|} \right\rceil$$

$$= a_i(1 - \frac{1}{|opt|}) \qquad (5)$$

Solving it, we get the following bound:

$$a_{i+1} \leq a_0(1 - \frac{1}{|opt|})^{i+1} \qquad (6)$$

Setting $i+1 = |opt| \cdot \ln \frac{a_0}{|opt|}$, we have:

$$a_{i+1} \leq a_0(1 - \frac{1}{|opt|})^{i+1}$$

$$= a_0(1 - \frac{1}{|opt|})^{|opt| \cdot \ln \frac{a_0}{|opt|}}$$

$$\leq a_0(\frac{1}{e})^{\ln \frac{a_0}{|opt|}} \qquad (7)$$

$$= a_0 \cdot \frac{|opt|}{a_0}$$

$$= |opt|$$

That is, after $|opt| \cdot \ln \frac{a_0}{|opt|}$ iterations, We only need to select at most $|opt|$ additional nodes to dominate the remaining nodes. The total number of nodes that we choose is no more than $|opt|$. Thus, the total number of dominators is $|opt| \cdot \ln \frac{a_0}{|opt|} + |opt|$.

Since $|opt| \geq \frac{n-2}{\Delta-1}$, the solution found by our algorithm has $|CDS|$ nodes,

$$|CDS| \leq |opt| + |opt| \cdot \ln \frac{a_0}{|opt|}$$

$$= |opt|(1 + \ln \frac{a_0}{|opt|})$$

$$\leq |opt| * (1 + \ln \frac{n(\Delta-1)}{n-2}) \quad (n \gg 2)$$

$$= |opt| * (\ln(\Delta-1) + 1) \qquad (8)$$

Therefore, the size of the CDS found by our algorithm is at most $(\ln(\Delta-1)+1)*|opt|$.

According to Theorem 2, the time complexity of EDC-DS algorithm is $O((e+n+\log n)\log_\Delta n)$. In step 3, we use Dijkstra algorithm to compute the shortest path between any two nodes of EDC-DS. This step needs $O(n\log(n+e))$ time. Thus, the time complexity of EDC-CDS algorithm is $O((e+n+\log n)\log_\Delta n + n\log(n+e))$.

# 5. Simulation

This section shows the simulation results and evaluates the performances of our DS and CDS construction algorithms. All simulations are implemented in C++. Here, we simulate the sizes of DS and CDS found by our algorithms under two scenarios with different transmission range. Simulation scenarios are as follows. A given number of nodes (ranging from 10 to 100 with a step of 10) are randomly distributed in a $100 \times 100$ space. Each node has a fixed uniform transmission range $r$ ($r$ is 25 and 50, respectively). There is no consideration for movement and channel collision of nodes. Thus, a pair of nodes are neighbors when their distance is no more than $r$. For each fixed number of nodes, we perform the simulation for 200 times and compute the average value.

Fig. 7 shows a sample network with 100 nodes whose transmission ranges are uniform, 25, where blue dots denote the nodes of the network, and the green edges denote the links of the network. If the distance between two blue dots, then there is a green edge to connect them. Links between nodes would change as the transmission radii of nodes change. Fig. 8(a) and 8(b) show the results obtained by our algorithms for constructing a DS and CDS, respectively. Where the bright nodes denote the dominators, and the dark nodes denote the dominated nodes.

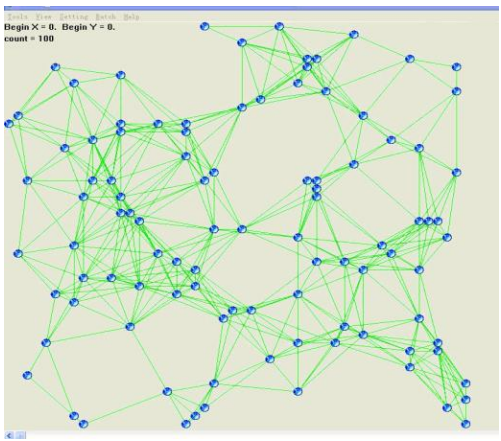

Fig. 7 A network topology: n=100, r=25.

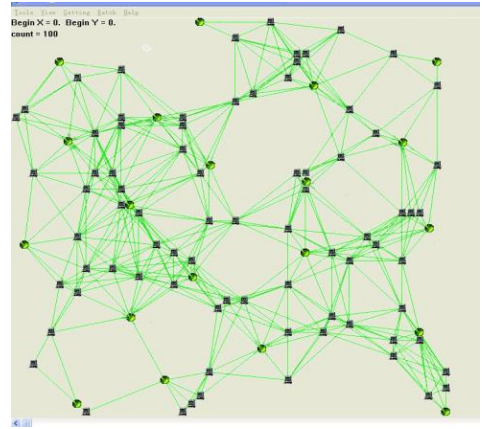

Fig. 8 (a) The DS constructed by our algorithm (n=100, r=25).

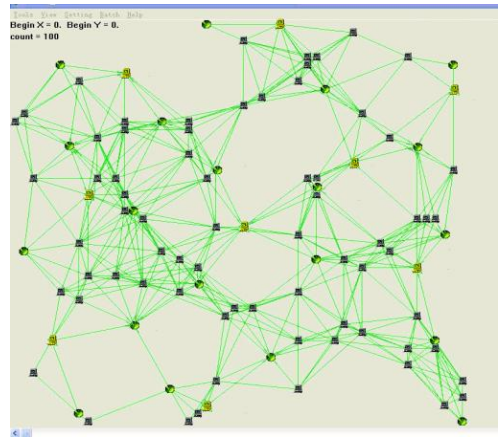

Fig. 8 (b) The CDS constructed by our algorithm (n=100, r=25).

Fig. 9(a) and 9(b) show the simulation results of the sizes of DSs when the transmission radii of nodes are 25. We compare our EDC-DS algorithm with MDS algorithm and WMDS algorithm [10]. Fig. 9(a) shows the trends when the number of nodes in the network ranges from 10 to 100 (the transmission ranges of the nodes are 25). From the curves, we can see that the number of dominators of EDC-DS algorithm is increasing as the network size increases. However, when the network size reaches a degree (n=100), the size of the DS is un-increasing or even reducing. This reason is that the network density is increasing as the network size increases, then each dominator can dominate more neighbors. Therefore, when the network density reaches a degree, the number of dominators is un-increasing. As depicted in Fig. 9(a), each algorithm has a good performance. In addition, when the total numbers of the nodes in the network are the same, the sizes of the DS of MDS algorithm and WMDS algorithm are larger than that of our EDC-DS algorithm. Fig. 9(b) shows the trends when the number of nodes in the network ranges from 10 to 100 (the transmission ranges of the nodes are 50). The performance of WMDS algorithm is still the worst, and our EDC-DS algorithm is the best. From Fig. 9(a) and 9(b), we can see that our approximate algorithm for DS in a dense network (the transmission range is 50) accounts for a smaller proportion of the total number of nodes than that of a sparse network (the transmission range is 50).

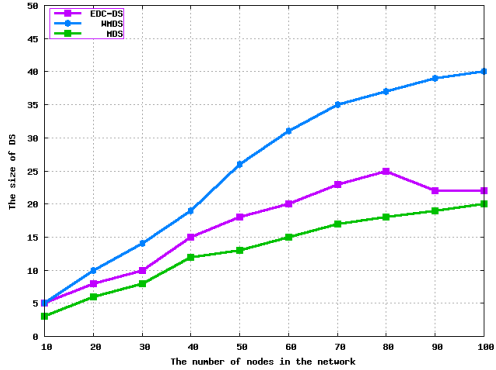

Fig. 9 (a) The number of dominators of EDC-DS (R=25).

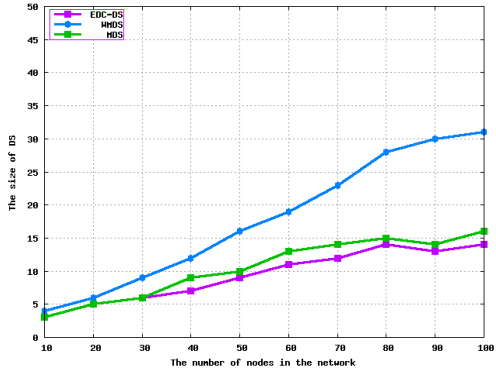

Fig. 9 (b) The number of dominators of EDC-DS (R=50).

Fig. 10(a) and 10(b) show the simulation results of the size of CDSs when the transmission radii of nodes are 25 and 50. We compare EDC-CDS algorithm with Wu's algorithm [23] and Das's algorithm [24]. Fig.10(a) shows the trends when the number of nodes in the network ranges from 10 to 100 and the transmission ranges of the nodes are 25. As shown in Fig. 10(a), when the total numbers of the nodes in the network are the same, the size of the CDS obtained by our algorithm is less than those of Da's algorithm and Wu's algorithm. Fig. 10(b) shows the trends when the number of nodes in the network ranges from 10 to 100 (the transmission ranges of the nodes are 50). Our algorithm is still the best. From Fig. 10(a) and 10(b), we can see that our approximate algorithm for CDS in a dense network with the transmission range being 50 accounts for a smaller proportion of the total number of nodes than that of a sparse network with the transmission range being 25.

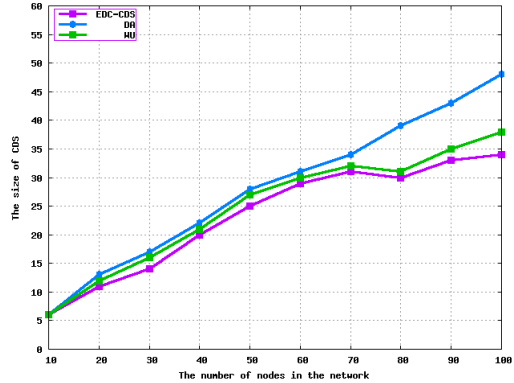

Fig. 10 (a) The number of dominators of EDC-CDS (R=25).

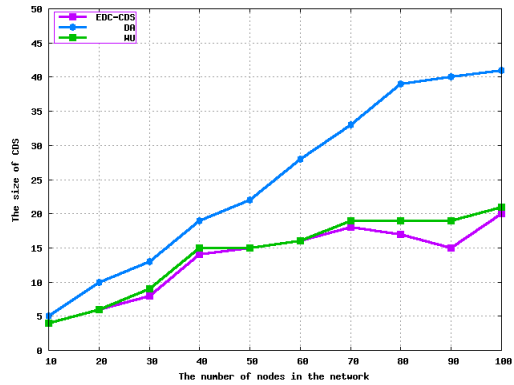

Fig. 10 (b) The number of dominators of EDC-CDS (R=50).

## 6. Conclusion and Future Work

This paper presents two novel approximate algorithms EDC-DS and EDC-CDS for DS and CDS, respectively. Both of the algorithms are based on edge dominating capability which was described in section 3. We evaluate the performance of the algorithms by theoretical analysis and simulation. According to the simulation results, we know that each of the proposed algorithms has a good performance to construct a DS or a CDS, especially in a dense network. In the future, we will develop other distributed algorithms based on edge dominating capability with better performance to solve the two problems. We also have interests on interference-aware algorithms based on edge dominating capability for DSs and CDSs.

**Acknowledgments**

This work was partially supported by the National Natural Science Foundation of China for contract (60373012, 11101243), Natural Science Foundation of Shandong Province for contract (ZR2012FM023, ZR2009GM009, ZR2009AM013), STPU of Shandong Province for contract (J10LG09) and TKP of Shandong Province for contract (2009GG10001014).

**Congcong Chen** received her B.S. degree in computer science and technology from Qufu Normal University, Shandong, China, in 2010. She is now a postgraduate in the School of Computer Science, Qufu Normal University. Her main research interest is wireless ad hoc and sensor networks.

**Jiguo Yu** received his Ph.D. degree in operational research and control theory from Shandong University, Shandong, China, in 2004. From 2007. He has been a full professor in the School of Computer Science, Qufu Normal University, Shandong, China. His main research interests include wireless networks, algorithms, peer-to-peer computing and graph theory. In particular, he is interested in designing and analyzing algorithms for many computationally hard problems in computer networks. He is a member of the IEEE, and a senior member of the CCF (China Computer Federation).

**Xiujuan Zhang** received her M.S. degree in computer science from Huazhong University of Science and Technology, Hubei, China, in 2004. She is now a lecture in the School of Computer Science, Qufu Normal University. Her main research interest is wireless ad hoc and sensor networks.